\documentstyle[prl,aps,multicol,ifthen]{revtex}
\newcommand{\showfigures}{no}     
\newcommand{\wide}[2]{
\end{multicols}
\widetext
\noindent
\ifthenelse{\equal{#1}{t}}
{}
{
\raisebox{0.1in}[0in][0.02in]{$\rule{3.575in}{0.002in}
\rule{0.002in}{0.08in}$}
}
#2
\ifthenelse{\equal{#1}{b}}
{}
{
{\raisebox{-0.1in}[0in][0.02in]
{\hspace{3.575in}$\rule{0.002in}{0.08in}
\rule[0.08in]{3.575in}{0.002in}$}
}
}
\begin{multicols}{2}
\noindent
}
\tighten
\ifthenelse{\equal{\showfigures}{yes}}{\input psfig}{} 
\begin{document}
\draft
\title{Role of Divergence of  Classical Trajectories in Quantum Chaos}
\author{ I.L. Aleiner$^{1,*}$ and A.I. Larkin$^{1,2}$}  
\address{
$^1$Theoretical Physics Institute, University of Minnesota, 
 Minneapolis, MN 55455 \\
$^2$L.D. Landau Institute for Theoretical Physics, 117940 Moscow, Russia}
\maketitle
\begin{abstract}
 We study  logarithmical in $\hbar$ effects in the
statistical description of quantum chaos. We found analytical
expressions for the deviations from the universality in the weak
localization correction and in the level statistics and showed that the
characteristic scale for these deviations is the Ehrenfest time
$t_E = \lambda^{-1} |\ln\hbar|$, where $\lambda$ is the Lyapunov exponent
of the classical motion. 
\end{abstract}

\begin{multicols}{2} 
It is accepted in literature to call the consideration of quantum
phenomena in classically chaotic system the ``quantum
chaos''\cite{chaos}. For the de~Broglie wavelength $\lambda_F$  much
smaller than the characteristic size of the system, the quantum phenomena
still bear essential features of the  classical chaotic motion.
Examples of such systems studied both theoretically and experimentally
are ballistic cavities or antidot arrays\cite{Baranreview}.
The quantities usually considered include different correlators of 
quantum spectra of the system (level statistics) as well as  of 
different response functions, {\em e.g.} fluctuations of the
conductance (mesoscopics) or the quantum correction to the averaged
transport coefficients (weak localization).

In principle, all the aforementioned characteristics can be found by
solving the one particle  Schr\"{o}dinger equation for the given system.
However, the Schr\"{o}dinger equation for such systems can not be solved
analytically. Substantial progress can be achieved in the statistical
approach to the  quantum chaos. In such approach one gives up attempts
to find a contribution of a single quantum state but instead studies
correlators averaged over large number of quantum states. The averaging
for a given system can be performed  either over wide range of energies
or over the applied magnetic field.

In the present Communication we apply the supersymmetry
description\cite{Efetov83,Muzykantskii95,Andreev96}   to investigate how
the universality is established in the statistical properties of the
system at  low frequencies. We will show that the  time  it takes to
establish the universality is $t_E = \lambda^{-1} |\ln
\hbar|$, where $\lambda$ is the Lyapunov exponent of the classical
motion. We will  express   deviations from
the universality in the level statistics and in the weak localization
corrections in terms of the single
renormalization function\cite{Aleiner96}. Finally, we emphasize the
necessity of the finite reqularizator in the Perron-Frobenius operator
in obtaining  physical results for the quantum corrections.

Let us first discuss the physical origin of the logarithmic in $\hbar$
corrections. In the semiclassical approximation, each classical
trajectory corresponds to the quantum mechanical amplitude. Quantum
phenomena in the system originate from the interference of the different
amplitudes. After the averaging, the most of the interference
contributions vanish. Only ones which survive are the products which
contains the pairs of the coherent amplitudes. Such coherent amplitudes
are contributed by the segments of the same classical trajectories and
the resulting products are expressed in terms of the probabilities to
find such segments. The latter probabilities are found by solving the
classical equation of motion. 
Most usable quantities are the probability
where the initial $i$ and final $f$ states coincide 
${\bf n}_f={\bf n}_i,\ {\bf r}_f={\bf r}_i$
 [we will denote this probability as ${\cal D_+}(t;
{\bf n}_i,\ {\bf r}_i)$]
or related to each other by time inversion 
${\bf n}_f=-{\bf n}_i,\ {\bf r}_f={\bf r}_i$  [we will denote this probability
as ${\cal D_-}(t; {\bf n}_i,\ {\bf r}_i)$]. Here
${\bf r}$ and ${\bf n}$ are the coordinate of the particle and the direction
of its momentum respectively. First quantity is relevant for the leading
approximation for the two point correlator of the density of
states (DoS)\cite{Berry85,Altshuler86,Andreev96,Bogomolny96}, whereas, the
second quantity is important for the weak localization correction to the
conductivity\cite{Aleiner96} and for the higher order approximations  for the
correlator of DoS, see below.

In what follows we will consider only ergodic systems. This means that
after some time the particle visits all the phase space allowed by the
energy conservation, i.e. the classical probabilities
${\cal D}_\pm$ averaged over  the  conditions 
${\bf n}_i,\ {\bf r}_i$ cease to 
depend on time and take  value of $1/S$, where $S$
is the volume of the system. It is very crucial  that
the equilibration time  for 
${\cal 
D}_-$ is parametrically larger than  that for the probability  ${\cal D}_+$.

Characteristic relaxation time for the probability  ${\cal D}_+$ is of
the order of the flying time of a particle  across the system
$\tau_{fl}\simeq L/v_F$ for the ballistic regime or the Thouless time
$\tau_T \simeq L^2/D$ for the diffusive regime, ($L$ is size of the
system, $D$ is the diffusion coefficient, and $v_F$ is the Fermi
velocity).    

On the other hand, in the strictly classical limit probability to have
${\bf n}_f=-{\bf n}_i,\ {\bf r}_f={\bf r}_i$ 
vanishes no matter how large traveling
time
$t$ is. This is due to the fact that the final state can be reached by
moving along a classical trajectory which coincides with the initial
one. It means that a particle must be reflected exactly backwards
from an obstacle. For the chaotic system, the measure for
such process is zero and that is why  ${\cal D_-}(t;
{\bf n}_i,\ {\bf r}_i)=0$. The 
only reason for this probability not to vanish is
that the initial and final conditions can not be specified with the
accuracy better than it is allowed by the uncertainty principle. Due to
this principle, the difference $|{\bf n}_f \times {\bf n}_i|=\delta
\phi_0$ can not be smaller than the diffraction spreading $\delta
\phi_0 \gtrsim\sqrt{\lambda_F/a}$, with
$a\gg\lambda_F$ being the characteristic spatial scale of the static
potential the particle moves in. In order to find the probability for
such close (but not coinciding)
${\bf r}_f,\ {\bf r}_i$, one has to take into  
account the fact that the motion of
the particle at the initial and final stages are correlated. This is
because the trajectory along which   the particle moves on the final
stage,
$\left[{\bf r}(t-t_1), - {\bf n}(t-t_1)\right]$ almost coincides with the
trajectory particle moved along at the initial stage, $\left[{\bf
r}(t_1), {\bf n}(t_1)\right]$. This problem is equivalent to the
consideration of the divergence of two classical trajectories (``1'' and
``2'') which start from the same point ${\bf r}_i$ with a small difference in
the directions of their momenta  $| {\bf n}_2(0) 
\times {\bf n}_1(0)|=\delta
\phi_0$, (it can be seen by the time inversion on the final  segment). In
the chaotic system the difference
$\delta\phi_{12}(t)=|{\bf n}_2(t)\times {\bf n}_1(t)|$
grows exponentially with time $\delta\phi(t)\simeq\delta\phi_0e^{\lambda
t}$ where $\lambda$ is the Lyapunov exponent of the classical chaotic
motion. Therefore, we have also for the given trajectory 
$\delta\phi(t_1)=|{\bf n}(t-t_1) \times {\bf
n}(t_1)|\simeq\delta\phi_0e^{\lambda t_1}$. In order to close the
trajectory at some time $t_1^* < t/2$, angle $\delta\phi(t_1^*)$ should
become of the order of unity and thus $t \gtrsim
(2/\lambda)\ln\left(1/\delta\phi_0\right)$. Taking into account 
$\delta\phi_0 \gtrsim\sqrt{\lambda_F/a}$ we conclude that the time it
takes to establish the equilibrium value of function ${\cal D_-}$ is
the Ehrenfest time
\begin{equation}
t_E=\frac{1}{\lambda}\ln\left(\frac{a}{\lambda_F}\right),
\label{tE}
\end{equation}
and ${\cal D}_-$ vanishes at smaller time, ${\cal D}_-\simeq \theta
\left(t-t_E\right)$.

The above discussion  leads us to the following expressions for the
Fourier transform of the classical probabilities ${\cal D}_\pm$ for the
frequencies $\omega$ smaller than inverse time of the travel of the
particle across the system
\begin{equation}
{\cal D}_+(\omega) = \frac{1}{S}\frac{1}{-i\omega^+};\quad
{\cal D}_-(\omega) = \frac{1}{S}\frac{\Gamma(\omega )}{-i\omega^+},
\label{Ds}
\end{equation}
where $\omega^+=\omega+i0$.
The denominators in Eqs.~(\ref{Ds}) reflects the ergodicity of the
system at large time and the renormalization function $\Gamma(\omega)$
describes the delay of ${\cal D_-}(t)$  with respect to ${\cal
D_+}(t)$ by the Ehrenfest time $t_E$
\begin{equation}
\Gamma (\omega) = \exp\left(i\omega t_E -\frac{\omega^2 \lambda_2
t_E}{\lambda^2}\right).
\label{Gamma}
\end{equation}
The second factor in Eq.~(\ref{Gamma}) characterizes the fluctuations of
the Lyapunov exponent and the parameter $\lambda_2$ is of the order of
$\lambda$. More details about the derivation of function $\Gamma$ can
be found in Ref.\cite{Aleiner96}. Appearance of the 
new time scale $t_E$ is the qualitative difference between quantum
chaos $a \gg \lambda_F$ and quantum disorder $a \lesssim \lambda_F$
regimes. (In the systems where the scale of the potential $a$ is different
from the transport mean free path $l_{tr}$, the criteria for quantum chaos
is $a \gg \sqrt{l_{tr}\lambda_F}$, see Ref.\cite{Aleiner96}).

A powerful method for the calculation of the averaged quantities is the
supersymmetric nonlinear $\sigma$-model pioneered by
Efetov\cite{Efetov83} for the disordered systems. Recently, 
the supersymmetric action was suggested by  by Muzykantskii and
Khmelnitskii\cite{Muzykantskii95} and more recently by Andreev {\em et.
al.}\cite{Andreev96} for the system in the ballistic regime. 
Effective action in Ref.~\cite{Andreev96} is defined by means of the
classical  Perron-Frobenius operator which differs from the first order
Liouville operator by the regularizator of the second order. This
approach enables one to perform the systematic semiclassical expansion for
the averaged quantities and to understand how the underlying classical
dynamics shows up in various quantum correction.

Partition function ${\cal Z}$ in the supersymmetry approach is given by
the functional integral\cite{Andreev96}
\begin{equation}
{\cal Z}\left\{J\right\}=
\int {D} Q(1) \exp\left[-\frac{\pi\nu}{2}\int d1 STr\left({\cal L}
+{\cal L}_J\right)\right],
\label{Z}
\end{equation}
with Lagrangians ${\cal L},\ {\cal L}_J$ being defined as  
\begin{eqnarray}
&&{\cal L}=\frac{i\omega^+}{2}\Lambda Q + T^{-1}\Lambda \hat{L} T
+ \frac{1}{4\tau}\left(\frac{\partial Q}{\partial\phi_1}\right)^2,
\quad Q=  T^{-1}\Lambda T
\nonumber\\
&&{\cal L}_J= i Q(1)\left[J_1 k \Lambda + \left( J_2(1)\Lambda_++
J_2\frac{k+1}{2}(\bar{1})\Lambda_-
\right)\right]
\label{action}
\end{eqnarray}
where $\nu$ is the density of states per unit area. We used the
short-hand notation
$1 = ({\bf n}_1, {\bf r}_1),\
\bar{1} = (-{\bf n}_1, {\bf r}_1),\ d1=d{\bf n}_1d{\bf r}_1/2\pi$, 
coordinate  ${\bf r}$ and the direction of the momentum ${\bf n}=
(\cos\phi, \sin\phi)$ characterize the position of the particle on the
energy shell, (we will restrict ourselves to two dimensional systems).
Liouvillean operator $\hat{L}$ describes the classical evolution on the
energy shell and defined by the Poisson bracket $\hat{L}\ \cdot=
\left\{\cdot,\ {\cal H}\right\}$, where ${\cal H}$ is the Hamiltonian
function. Last term in the Lagrangian ${\cal L}$ is the regularizator
physical significance of which will be discussed later. Operation of
supertrace is defined in Ref.\cite{Efetov83}. We will consider only
systems with the unbroken timereversal symmetry (orthogonal ensemble).

In Eq.~(\ref{action}), $\hat{T}$ is $8 \times 8$ supermatrix  defined in
a linear superspace
$p\otimes g\otimes d$ which we represent as the direct product of three
linear spaces; $p$ and $d$ are the spaces of retarded-advanced and 
time reversal (complex conjugate) $2
\times 2$ matrices respectively, and
$g$ is the superspace of fermion-boson $2 \times 2$ supermatrices.
All the relevant matrices can be conveniently expressed in terms of the
Pauli matrices
$\tau^\alpha_z,\ \tau^\alpha_\pm=(\tau^\alpha_x \pm i \tau^\alpha_y)/2$,
 acting in spaces $\alpha=p,g,d$. Matrices $\Lambda = \tau^p_z
\otimes 1^g
\otimes 1^d$, $\Lambda_\pm = \tau^p_\pm \otimes 1^g \otimes
1^d$  and $k = 1^p \otimes \tau_z^g \otimes
1^d$  break the symmetry in advanced-retarded and in fermion-boson
spaces respectively. Matrix $T$ is the subject to constraints
$T^\dagger K T = K$ and $T^\dagger(1) = CT^T(\bar{1})C^T$,
where $2K=\left[({1+\tau_z^p})\otimes 1^g
+ ({1-\tau_z^p})\otimes \tau_z^g
\right] \otimes 1^d$
 and matrix of charge conjugate $C$ is given by $C = 1^p\otimes
\left(1^g\otimes\tau^d_- - \tau^g_z\otimes\tau^d_+\right)$. 

Partition function (\ref{Z}) allows to find different quantum mechanical
correlators averaged over the wide range of energies. Two point
correlator of the DoS, $R(\omega)=
\Delta^2\langle
\rho(\epsilon+\omega)\rho(\epsilon)\rangle_\epsilon$ (where 
$\rho(\epsilon)=Tr\delta(\epsilon - \hat{H})$, with $\hat{H}$ being the
Hamiltonian of the system, and
$\Delta$ is the mean level spacing ) and two-particle Green function
${\cal D}$ can be found as certain derivatives of action
\begin{mathletters}
\begin{eqnarray}
R(\omega)= -\frac{\Delta^2}{16\pi^2}\Re \left.
\frac{\partial^2Z}{\partial^2J_1} \right|_{J_{1,2}=0},
\label{R}\\
{\cal D}(\omega; 1, 2) = -
\frac{2\pi}{\nu}\left.
\frac{\delta^2Z}{\delta J_2(1)\delta J_2(\bar{2})}
\right|_{J_{1,2}=0}.
\label{D}
\end{eqnarray}
\end{mathletters}

We will be interested in effects associated with the energy scale
$\omega \simeq t_E^{-1} \gg \Delta$, and will not consider contributions
oscillating on $\Delta$. For analysis in this situation it suffices to
consider only small fluctuations of $Q$ around
$\Lambda$. We use the standard parametrization
\begin{equation}
\!T\!\!=\!\!\frac{1\!+\!iP}{\sqrt{1\!+\!P^2}},\ 
P\Lambda\! =\!-\Lambda P\!,\ 
KP^\dagger\! K\!\! =\!P\!, \  P(1)\!=\!-\bar{P}(\bar{1}),
\label{parametrization}
\end{equation}
where the operation of time reversal is defined as
\begin{equation}
\bar{M}= KCM^TC^TK,
\label{timereversal}
\end{equation}
for arbitrary 
supermatrix  $M$.

Substituting formula Eq.~(\ref{parametrization}) into
Eq.~(\ref{action}) and keeping terms up to the fourth order in $P$ we
obtain ${\cal L} = {\cal L}_0 + {\cal L}_{int}$ where 
the quadratic part of the Lagrangian describing the classical
motion is given by
\begin{equation}
{\cal L}_0=P(-i\omega + \hat{L}_R)P, \quad
\hat{L}_R=\hat{L}-\frac{1}{\tau}\frac{\partial^2}{\partial\phi^2}.
\label{quadratic}
\end{equation}
Operator $\hat{L}_R$ is known as the Perron-Frobenius operator.
Quartic part, responsible for lowest order quantum corrections
has the form
\begin{equation}
{\cal L}_{int}=-P^3(-i\omega+ \hat{L}_R)P + \frac{1}{\tau}
\left(P\frac{\partial P}{\partial\phi}\right)^2.
\label{quartic}
\end{equation}

We expand the partition function up to the first order in ${\cal
L}_{int}$ and up to the second order in the source Lagrangian
\[
{\cal L}_J=iJ_1k\left(1-2P^2+2P^4\right)
+2J_2\left( k+1 \right)\Lambda_+\left( P - P^3\right),
\]
where we used Eq.~(\ref{parametrization}) and omitted traceless terms.
 For  calculating
the averages of the arising products of matrices $P$ we use the Wick
theorem with the  contraction rules
\wide{m}{
\begin{eqnarray}
&&2\pi\nu P\!\!\overbrace{\ (1) M\ }\!\!P(2)=
{\cal D}^0(\bar{1}, \bar{2})\Lambda_\parallel^+
STr \left[M\Lambda_\parallel^- \right]+
{\cal D}^0(1, 2)\Lambda_\parallel^-
STr\left[ M\Lambda_\parallel^+\right] +
{\cal D}^0(1, \bar{2})
\Lambda_\parallel^- \bar{M}\Lambda_\parallel^+ +
{\cal D}^0(\bar{1}, {2})
\Lambda_\parallel^+ \bar{M}\Lambda_\parallel^-,
\label{contraction}\\
&&2\pi\nu STr [MP\!\!\overbrace{\ (1)] \ STr [N\ }\!\!P(2)]=
STr \left[\left({\cal D}^0(\bar{1}, \bar{2})
M
- {\cal D}^0(\bar{1}, 2)
\bar{M}\right)\Lambda_\parallel^-
N\Lambda_\parallel^+ +
\left({\cal D}^0(1, 2)
M
 -
{\cal D}^0(1, \bar{2})
\bar{M}\right)\Lambda_\parallel^+
N\Lambda_\parallel^-\right],
\nonumber
\end{eqnarray}
}
\noindent
where matrices $\Lambda_\parallel^\pm =
\left(1\pm\Lambda\right)/2$ break the symmetry in the retarded-advanced
subspace, and the classical propagator
${\cal D}^0$ is the solution of the equation
\begin{equation}
{\left(-i\omega^+ + \hat{L}_R\right)_1}{\cal D}^0(1,2) = 
{2\pi}\delta_{12}.
\label{propagator}
\end{equation}

Substituting the result of the averaging in Eq.~(\ref{D}), we obtain
${\cal D}(1,2)={\cal D}^0(1,2) + \delta {\cal D}(1,2)$, where the
quantum correction to the classical propagator is given by
\begin{eqnarray} 
&&{2\pi\nu}\delta{\cal D}(1,2) =  {\cal D}^0(1,\bar{2})
{{\cal D}^0(\bar{2},2)} +
{{\cal D}^0(1,\bar{1})}{\cal D}^0(\bar{1},2) 
\label{WLC}\\ &&+\int d3 
 {\cal D}^0(1,3){\cal D}^0(\bar{3},2)
\left( 2i\omega-
\hat{L}_R\right)_3
{{\cal D}^0(3,\bar{3})}.
\nonumber
\end{eqnarray}
Notice, that Eq.~(\ref{WLC}) gives a correction only to the
non-zero modes of the Perron-Frobenius operator, $\int d1\delta {\cal
D}(1,2)=0$, which is a consequence of the charge conservation.

Analogously, we obtain with the help of Eq.~(\ref{R}) the
following expression for the two-point correlator of DoS,
$R=R^0 + \delta R$. Here 
\begin{mathletters}
\label{statistics}
\begin{equation}
R^0(\omega)=1+\frac{\Delta^2}{\pi^2}{\rm Im} \int d1 
\frac{\partial {\cal D}^0(1,{1})}{\partial \omega},
\label{statistics1}
\end{equation}
which is well-known result for the disordered\cite{Altshuler86}
and chaotic systems\cite{Andreev96}. Quantum correction, obtained for the
first time in the present paper, has the form
\begin{eqnarray}
\delta R(\omega)=\frac{\Delta^2}{2\pi^3\nu}
{\rm Re}
\left[4\pi\nu\int\!\! d1d2 {\cal D}^0(1,{2})\delta{\cal D}^0(2,1)
-
\right.\nonumber\\ 
\left.
\int\!\! d3 
\frac{\partial
{\cal D}^0(\bar{3}, 3)}{\partial\omega}
\left( 2i\omega-
\hat{L}_R\right)_3
\frac{\partial{\cal D}^0(3 ,\bar{3})}{\partial\omega}
\right].
\label{statistics2}
\end{eqnarray}
Operator $\hat{L}_R$  in Eqs.~(\ref{WLC}) and (\ref{statistics2}) acts
on both arguments $3,\bar{3}$. Deriving Eqs.~(\ref{statistics}), we
used identity $-i\partial_\omega {\cal D}^0(1,2) = \int d3
{\cal D}^0(1,3){\cal D}^0(3,2)$.
\end{mathletters}

Equations (\ref{WLC}) and (\ref{statistics2}) describe 
the lowest quantum
corrections expressed in terms of the
solutions of the  Liouville equation (with the regularizator added) for
a given system, where no ensemble averaging is assumed. They are both
determined by the classical propagators between the points related by the
time inversion. As we already explained, the equilibration time for such
probabilities is the parametrically large Ehrenfest time.

Let us first discuss the two-point DoS correlator $R$. The classical
propagator ${\cal D}^0$ entering  into Eq.~(\ref{statistics1}) is the
probability  of the return to the initial state and at energies
$\omega$ much smaller than the Thouless energy it coincides with ${\cal
D}_+$ from Eq.~(\ref{Ds}). 
First term in Eq.~(\ref{statistics2}) correspond to the weak
localization corrections of non-zero modes of Perron-Frobenius operator and it
can be neglected at such low-frequencies. Propagators entering into the second
term are the classical probabilities with the initial and final states
related by the time inversion and they coincide with ${\cal D}_-$ at low
frequencies. Substituting Eqs.~(\ref{Ds}) into Eqs.~(\ref{statistics}),
we obtain
\begin{equation}
R(\omega) = 1- \frac{\Delta^2}{\pi^2\omega^2}
+ \frac{\Delta^3\omega}{\pi^3}{\rm Im}
\left(\frac{\partial}{\partial\omega}\frac{\Gamma
\left(\omega\right)}{\omega}\right)^2 + \dots,
\label{rfinal}
\end{equation}
where the renormalization function $\Gamma$
defined in Eq.~(\ref{Gamma}). It is easy to see that result
(\ref{rfinal}) does not contain terms linear in $t_E$. Actually, this can
be proven for all orders of perturbation theory in $\Delta/\omega$ by using
the approach similar to Ref.\cite{Mirlin94}. Even though the found correction
to the universal Dyson result\cite{book} for the orthogonal ensemble is
small, it oscillates with the period $t_E^{-1}$, where $\Delta \ll t_E^{-1}
\ll \tau_{fl}$ and, therefore, can be distinguished.

The result for the quantum correction
to the conductivity of infinite system $\delta\sigma$  obtained from
Eq.~(\ref{WLC}) 
\[
\delta \sigma (\omega ) = -  
\frac{e^2}{2\pi^2\hbar}\ln\left(\frac{1}{\omega\tau_{tr}}\right)
\Gamma^2\left(\omega\right).
\]
is renormalized in comparison with the quantum disorder
regime\cite{Gorkov79} by $\Gamma$.
Details of the derivation of this equation from
Eq.~(\ref{WLC})  can be found in Ref.~\cite{Aleiner96}.

Ehrenfest time (\ref{tE}) contains de~Broglie wavelength $\lambda_F$.
This scale is already absent in the effective $\sigma$-model
(\ref{action}) which is formulated on the Hilbert space of 
functions smooth on scale $\lambda_F$, and the lower cut-off of the
logarithm is related  to the reqularizator
$1/\tau$ in the Lagrangian (\ref{action}). It follows from the
solution of Eq.~(\ref{propagator}),\cite{Aleiner96} that ${\cal D}_-$
has the form (\ref{Ds}) with  
$t_E =\lambda^{-1}
\ln\left(\lambda\tau\right)$. Therefore, the regularizator can not be
put to zero even in the end of the calculation and it should be assigned
some physical value. 
The value of the reqularizator can be derived for the case if, in
addition to the semiclassical potential, there are also quantum
impurities in the system which provide the small angle
scattering. The diffraction on the semiclassical potential itself is
described by an equation more complicated than (\ref{propagator}) and
the regularizator was not found consistently within the
$\sigma$-model approach. However, it is not really necessary because the
dependence on the regularizator is only logarithmical. Using
Eq.~(\ref{tE}) obtained by different arguments, we conclude that the
physical value of the regularizator is given by
\begin{equation}
\frac{1}{\tau}= \lambda \frac{\lambda_F}{a}.
\label{physics}
\end{equation}

There is a subtlety in Eq.~(\ref{WLC}) which deserves more discussion
because it helps to understand the importance of the reqularizator in
the supersymmetric Lagrangian  (\ref{action}).
With the help of Eq.~(\ref{propagator}), one can rewrite
Eq.~(\ref{WLC}) to the more compact form
\[
\delta{\cal D}(1,2)= \int d3 \frac{{\cal
D}^0(3,\bar{3})} {\pi\nu\tau}
\frac{\partial {\cal D}^0(1,3)}{\partial\phi_3}
\frac{\partial {\cal D}^0(\bar{2},3)}{\partial\phi_3}.
\]
Naively, $\delta{\cal D}\to 0$ for $\tau\to\infty$. However, this
contribution is anomalous and caution should be exercised while taking
the limits. Namely,
${\cal D}^0(1,\phi_3+\delta\phi){\cal D}^0(2,\phi_3-\delta\phi)$ is a
singular function on $\delta\phi$ and this singularity is  cut-off
by the same regularizator $1/\tau$. As the result, the derivatives
over $\phi_s$ are proportional to $\sqrt{\tau}$ and the dependence on
the regularizator remains only logarithmical, see Eq.~(\ref{WLC}).

In conclusion, we studied  logarithmical in $\hbar$ effects in the
statistical description of quantum chaos. We found analytical
expressions for the deviations from the universality and showed that the
characteristic scale for these deviations is the Ehrenfest time
$t_E = \lambda^{-1} |\ln\hbar|$, where $\lambda$ is the Lyapunov exponent
of the classical motion. Finally, we discussed role of anomalies in the
supersymmetric $\sigma$-model\cite{Andreev96} of quantum chaos.

Interesting discussions with O. Agam and B.L. Altshuler are 
gratefully acknowledged. 

\end{multicols}
\end{document}